\documentclass[twocolumn,11pt]{article}
\usepackage{cite}

\usepackage{amsthm,amsmath,xfrac}
\usepackage{amssymb}
\usepackage{textcomp}
\usepackage{hyperref}
\usepackage{a4wide}
\usepackage[applemac]{inputenc} 
\usepackage{setspace}
\usepackage{verbatim}
\usepackage{placeins}

\newcommand{\degree}{$^{\circ}\,$}
\newcommand{\textapprox}{\raisebox{0.5ex}{\texttildelow}}

\usepackage{natbib}
\let\cite\citep

\title{Whole genome mapping of 5' RNA ends in bacteria by tagged sequencing :  A comprehensive view in  \textit{Enterococcus faecalis}}

\author{Nicolas Innocenti$^{1,2,3}$, Monica Golumbeanu$^{4,5}$, Aymeric Fouquier~d'H\'erou\"el$^{1,6}$, \\ Caroline Lacoux$^{2,3}$, R\'emy A. Bonnin$^{7}$, Sean P. Kennedy$^{8}$, Fran\c coise Wessner$^{2,3}$,\\ Pascale Serror$^{2,3}$, Philippe Bouloc$^{7}$, Francis Repoila$^{*2,3}$, Erik Aurell$^{*1,9}$}

\begin{document} 
\onecolumn
\maketitle

{\bf Short title : }  Whole genome mapping of 5' RNA ends in bacteria

{\bf E-mail : } njain@kth.se, monica.golumbeanu@bsse.ethz.ch, aymeric.dherouel@uni.lu, caroline.lacoux@jouy.inra.fr, remy.bonnin@u-psud.fr, sean.kennedy@jouy.inra.fr, \\ francoise.wessner@jouy.inra.fr, pascale.serror@jouy.inra.fr, philippe.bouloc@u-psud.fr,\\  {\bf francis.repoila@jouy.inra.fr}, {\bf eaurell@kth.se}. 

\begin{itemize}
\item[$^{1}$] Department of Computational Biology, KTH Royal Institute of Technology, AlbaNova University Center, Roslagstullsbacken 17, SE-10691 Stockholm, Sweden
\item[$^{2}$]  INRA, UMR1319 Micalis, Domaine de Vilvert, F-78352, Jouy-en-Josas, France
\item[$^{3}$]  AgroParisTech, UMR Micalis, Domaine de Vilvert, F-78350, Jouy-en-Josas, France
\item[$^{4}$]  Department of Biosystems Science and Engineering, ETH Z\"urich, Mattenstrasse 26, CH-4058, Basel, Switzerland
\item[$^{5}$]  SIB Swiss Institute of Bioinformatics, University of Basel, Klingelbergstrasse 50-70, CH-4056, Basel, Switzerland
\item[$^{6}$]  Luxembourg Centre for Systems Biomedicine, University of Luxembourg, 7, avenue des Hauts Fourneaux, L-4362, Belval, Luxembourg
\item[$^{7}$]  Institut de G\'en\'etique et Microbiologie, Universit\'e Paris-Sud, CNRS, UMR8621, 15, rue Georges Cl\'emenceau, F-91405, Orsay, France
\item[$^{8}$]  INRA, MetaGenoPolis US1367, Domaine de Vilvert, F-78350, Jouy-en-Josas, France
\item[$^{9}$]  Department of Information and Computer Science, Aalto University, Konemiehentie 2, FI-02150 Espoo, Finland
  \item[$^*$] Co-corresponding authors.
\end{itemize}


\clearpage
\twocolumn
\begin{abstract} 

\textit{Enterococcus faecalis} is the third cause of nosocomial infections. To obtain the first comprehensive view of transcriptional organizations in this bacterium, we used a modified RNA-seq approach enabling to discriminate primary from processed 5'RNA ends. We also validated our approach by confirming known features in \textit{Escherichia coli}.

We mapped 559 transcription start sites and 352 processing sites in \textit{E. faecalis}. A blind motif search retrieved canonical features of SigA- and SigN-dependent promoters preceding TSSs mapped. We discovered 95 novel putative regulatory RNAs, small- and antisense RNAs, and 72 transcriptional antisense organisations.

Presented data constitute a significant insight into bacterial RNA landscapes and a step towards the inference of regulatory processes at transcriptional and post-transcriptional levels in a comprehensive manner.
\end{abstract}

\section*{Introduction}
\textit{Enterococcus faecalis} is a ubiquitous Gram-positive bacterium and one of the first colonizers of the human gastro-intestinal tract after birth. It belongs to the core-microbiota and lives in the guts during the entire human life, suggesting a contribution of the bacterium to intestinal homeostasis \cite{Adlerberth2009,Campeotto2007,Qin2010}. In contrast to this potentially beneficial role, \textit{E. faecalis} is also the third cause of nosocomial infections and may carry and transfer various antibiotic resistances to other bacterial species, making its presence in the medical environment a serious concern \cite{Arias2012}. 
 The opportunism of \textit{E. faecalis}, i.e. the transition from commensalism to pathogenicity in response to environmental cues, underlines its capacity to adapt and survive to harsh conditions. 
  Thus, deciphering the regulatory pathways that enable \textit{E. faecalis} to undergo the transition from commensalism to pathogeny is a key component in the understanding the dual lifestyle of this microorganism \cite{Gilmore2003}. The V583 strain was one of the first discovered vancomycin-resistant clinical isolates of \textit{E. faecalis} \cite{Sahm1989}. Its genome, a circular chromosome (3\,218~kbp) and three circular plamids pTEF1 (66~kbp), pTEF2 (57.7~kbp) and pTEF3 (18~kbp), contains at least 3264 annotated protein-coding genes \cite{Paulsen2003}. Although partial transcriptomic analyses have been performed \cite{Aakra2010,Opsata2011,Vebo2009,Vebo2010}, a comprehensive and dynamic view of the RNA landscape of V583 is missing. \\
Whole-transcriptome studies of prokaryotes via tiling arrays and RNA sequencing (RNA-seq) have unveiled a plethora of actively transcribed RNAs, and highly complex transcriptional organizations due to numerous promoters nested in open reading frames (ORFs), antisense (asRNAs) and small RNAs (sRNAs) genes (among other reviews \cite{Georg2011,toledo-arana2010}). Although these global studies have been extremely valuable, their functional and regulatory insights remain incomplete as primary and processed RNAs cannot be distinguished and hence transcriptional (RNA synthesis) and post-transcriptional processes (RNA processing and stability) cannot be separated. The use of differential RNA-seq (dRNA-seq), an astute method that enriches an RNA population for primary transcripts, partially overcomes these limitations and gives access to the primary transcriptome \cite{Albrecht2010,Bohn2010,Irnov2010,Sharma2010}. Yet, a major limitation of dRNA-seq is that all transcripts cannot be detected in a single experiment as they are degraded by a 5'-phosphate-dependent exonuclease, and thus information on post-transcriptional events is lost \cite{Sharma2010}. Global scale analysis of RNA stability has been performed in a few bacterial species, e.g. \textit{Bacillus cereus} \cite{Kristoffersen2012}, \textit{Bacillus subtilis} \cite{Hambraeus2003}, \textit{Escherichia coli} \cite{Esquerre2013,Mohanty2006,Selinger2003}, \textit{Mycobaterium tuberculosis} \cite{Rustad2013}, \textit{Lactococcus lactis} \cite{Redon2005} and \textit{Prochlorococcus} \cite{Steglich2010}. These "stabilomes" have highlighted the broad and crucial contribution of RNA stability to gene expression reprogramming when bacteria face stresses, adapt to novel nutrient conditions or grow at different rates. Yet, for stabilomes, measurements consider transcribed regions as unique entities, where different sorts of RNA molecules can be present and cannot be seen. \\
We previously described a method that enables us to differentially tag 5' ends of primary and processed RNAs \cite{FouquierdHerouel2011}. In the present work, we have coupled this method to RNA-seq, yielding novel insights into the bacterial transcriptome landscape where the primary and the processed RNAs are unveiled within a single experiment; we call the totality of primary and processed RNAs the ppRNome. 
We have sorted transcription start sites (TSSs) and processing sites (PSSs) and validated the method by reproducing known results for \textit{E. coli}.  The presented data provide a first comprehensive transcriptional landscape of the human pathogen \textit{E. faecalis}.

\section*{Results and Discussion}

\subsection*{Global view of the \textit{E. faecalis} RNA landscape}
Bacterial native (or primary) transcripts undergo cleavage that can be maturation or degradation processes \cite{Rochat2013}. Without the ability to identify and discriminate primary from processed transcripts, we have only partial information of gene expression control at the genome scale. In bacteria, transcriptional start sites (TSSs, or "+1") are characterised by the presence of a 5' triphosphate group. In contrast, 5' RNA ends created by endonucleolytic cleavages (PSSs) are 5' monophosphate. We exploit this chemical difference by labelling differentially mono- and triphosphate 5' RNA ends with two short RNA oligonucleotides, the "tags" (Materials and Methods and Section S1 in supplementary data) \cite{FouquierdHerouel2011}. We have combined this differential 5' RNA end tagging with deep sequencing technologies and termed it "tagRNA-seq" to visualise the primary and processed transcripts of \textit{E. faecalis} in a comprehensive manner (Figure 1 and "The ppRNome browser" website, see section "Visualisation of Results" in Materials and Methods). TagRNA-seq was performed on total RNAs extracted from bacteria grown in static (S) and respiratory (R) conditions, providing transcriptomes coined "St" and "Rt", respectively (Section S2, Table S1). In parallel to these, and as control, three other RNA libraries from \textit{E. faecalis} and one from \textit{E. coli} were sequenced on different next generation sequencing platforms  (See Materials and Methods and Section S2). In order to account for variations in total number of reads and to be able to compare experiments, RNA levels are reported normalised to the total number of reads mapped, as commonly done in RNA-seq \cite{20196867}. Additionally, the ligation procedure introduces a new variability in the experiment that is corrected for by normalising the number of tagged reads mapped at a given position to the total number of tagged reads mapped for the entire V583 genome (Table SA). 

 \begin{figure*}[htbc]
  \centering
  \includegraphics[width=0.95\textwidth]{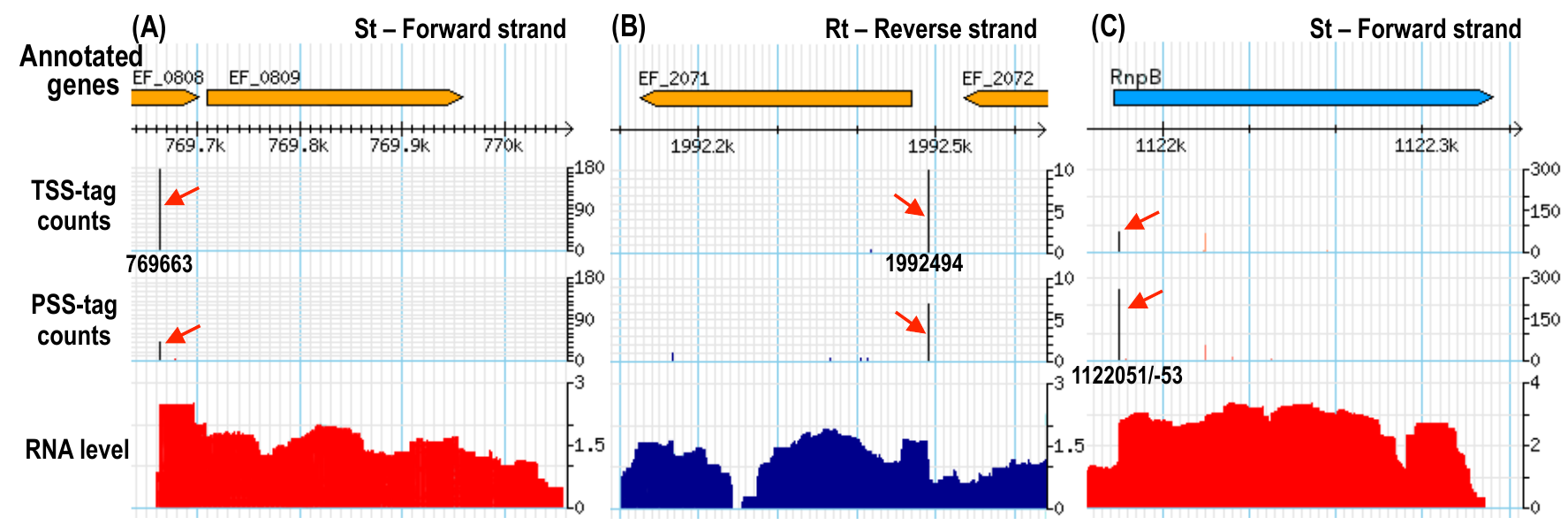}
  \caption{ \textbf{Three examples of 5' RNA ends viewed by the ppRNome on the \textit{E. faecalis} V583 chromosome from the "ppRNome" browser.} Below the line "Annotated genes", coordinates are those of the chromosome. The location of tags detected is indicated by the black vertical lines and the red arrows. TSS-tags are shown in the upper line, PSS-tags in the line below. "RNA levels" show the RNA signal detected; in red from St, in blue from Rt. Accurate values obtained for TSS- and PSS-tag counts and RNA levels are provided Table SA. {\bf (A)} Transcription start site mapped at 769663/-5 for \textit{ef0809}. This TSS could be easily predicted from the signal coverage. {\bf (B)} TSS mapped at 1992494 for \textit{ef2071}. This TSS is internal to the signal provided by the transcription of \textit{ef2072} and would be difficult to predict. {\bf (C)} Processing site mapped at 1121951/-53 for the RNA RnpB. This PSS is a dozen nucleotides downstream from the previously mapped TSS (see section "Processing sites").}
\end{figure*}

Globally, St and Rt show that significant transcription occurs in a limited portion of the \textit{E. faecalis} genome. Out of the \textapprox3.34 Mbp long genome, \textapprox1.65 Mbp appears to be transcribed in each condition (coverage greater than 2x), including \textapprox90 kbp due to antisense transcription and \textapprox470 kbp made up by non-annotated and/or non-coding portions, i.e. 5'- and 3' untranslated regions (UTRs), unannotated ORFs, and as- and sRNAs (see below). These data are in line with previous reports highlighting that the information provided by genomic annotations of bacterial genomes on their gene content remains incomplete \cite{Albrecht2010,Irnov2010,Mitschke2011,Sharma2010,Wurtzel2012}.

\subsection*{5' tagging of RNA ends: analysis and interpretation}
We compared deep sequencing data obtained with tagged and untagged RNA libraries prepared from \textit{E. faecalis} grown in S conditions. We then predicted 5' RNA ends by analysing edges of sequence coverage signals in the transcribed regions (Materials and Methods and Section S3). Predictions obtained from both RNA libraries show good agreement, indicating that the tagging procedure does not affect the location of transcription edges in the resulting coverage (Section S3, Figure S3 and the ppRNome browser). Moreover, several TSSs mapped previously by other methods were retrieved by tagRNA-seq at near-identical locations ($\pm$ 2 bp) attesting to the reliability of the method. For example, we find the TSSs of \textit{sodA} (\textit{ef0463}), coding for the superoxide dismutase, \textit{ptb}, coding for the phosphotransbutyrylase, \textit{fsrB/D} (\textit{ef1821}), coding for the cysteine protease-like processing enzyme FsrB and the autoinducing propeptide FsrD of the \textit{fsr} system, a homologue of the accessory gene regulator (\textit{agr}) of \textit{Staphylococcus aureus}, and \textit{gelE} (\textit{ef1818}), coding for a gelatinase \cite{Nakayama2006,Qin2000,Qin2001,Verneuil2006,Ward2000} (see below and Table SB). 

In the ideal case, the procedure should identify unambiguously TSSs and PSSs. In practice, a fraction of 5' ends attached to a TSS-tag were also ligated to a PSS-tag. Indeed, \textit{in vivo}, 5'-triphosphate RNA ends are enzymatically converted to monophosphate, often as a first step of RNA degradation \cite[and references therein]{Bail2009}. Therefore, a fraction of TSSs are expected to be associated with the PSS-tag. This effect may be further strengthened by spontaneous hydrolysis of 5'-triphosphate RNA ends during the ligation step of the PSS-adaptor and preceding RNA treatments, generating 5'ends opened for ligation. On the other hand, the first step of the tagging procedure using the T4 RNA ligase is certainly not complete and acts with different efficiency on  different RNA molecules \cite{Zhuang2012,Raabe2014}. Therefore, at the second ligation step, 5' monophosphate ends (i.e. PSSs) that have escaped the first tag can be ligated to the TSS-adaptor and appear as false TSSs. For each 5' RNA end mapped in this study, figure 2A presents the number of each tag counted. The distribution of 5' termini extends continuously between the two axes and hence does not give an immediate way to distinguish TSSs from PSSs. However, the distribution can be sorted by additional arguments, paying the price of discarding information on a fraction of mapped positions. 1) PSSs (i.e. 5' monophosphate groups) for which the first ligation step was partial and also tagged with the TSS-tag sequence at the second step, should not give more TSS-tag counts than PSS-tag counts since the enzyme should act with the same efficiency on the same RNA end at each step. Therefore, points (i.e. 5' RNA ends) above the diagonal may be either TSSs or partially ligated PSSs, but 5' RNA termini falling below the diagonal in Figure 2A should be  TSSs. Obviously, such a cutoff eliminates true TSSs that would exist \textit{in vivo} mainly as 5' monophosphate ends. 2) In accordance with the previous argument, all other TSSs known from the literature fall below the diagonal with one exception (Figure 2B), the ncRNA Ref25C (\underline{R}NA in \textit{\underline{E}. \underline{f}aecalis} 25C), which we discuss in more details in Section S4. 3) We considered separately 5' edges of transcribed regions that feature an absence of detectable expression upstream and should therefore be a signature of a TSS. As expected for those selected RNA ends, and in accordance with the two first arguments, a clear distribution below the diagonal appears (Figure 2C). 4) A motif search in DNA regions upstream 5' RNA ends located below the diagonal shows that more than 80\% of them contain at least one canonical sequence featuring a promoter region ($-$10 and/or $-$35 boxes). In contrast, the same search performed for 5' RNA ends above the diagonal does not retrieve any sequence reminiscent of a canonical promoter region (see below). The presence of promoter motifs in one area delineated by the diagonal is a very strong argument in favor of the location of true TSSs below 45\degree in the plot presented in Figure 2A. Considering these rules and in order to err on the side of caution, in this work we will only consider points (i.e. 5' RNA ends) below 30\degree as "TSSs", and above 60\degree as "PSSs"; for points in between, 5' RNA ends cannot be assigned with certainty and will be considered as undetermined. Compared to other single nucleotide resolution RNA-seq methods, tagRNA-seq provides for the first time, an accurate mapping of TSSs buried in transcribed regions and of RNA cleavage sites at a comprehensive scale in a single view, without requiring comparison between transcriptomes \cite{Nicolas2012,Sharma2010,Wurtzel2012}, (Figure 1, the "ppRNome" browser and Table SA).

\begin{figure}[!h]
  \centering
  \includegraphics[width=0.47\textwidth]{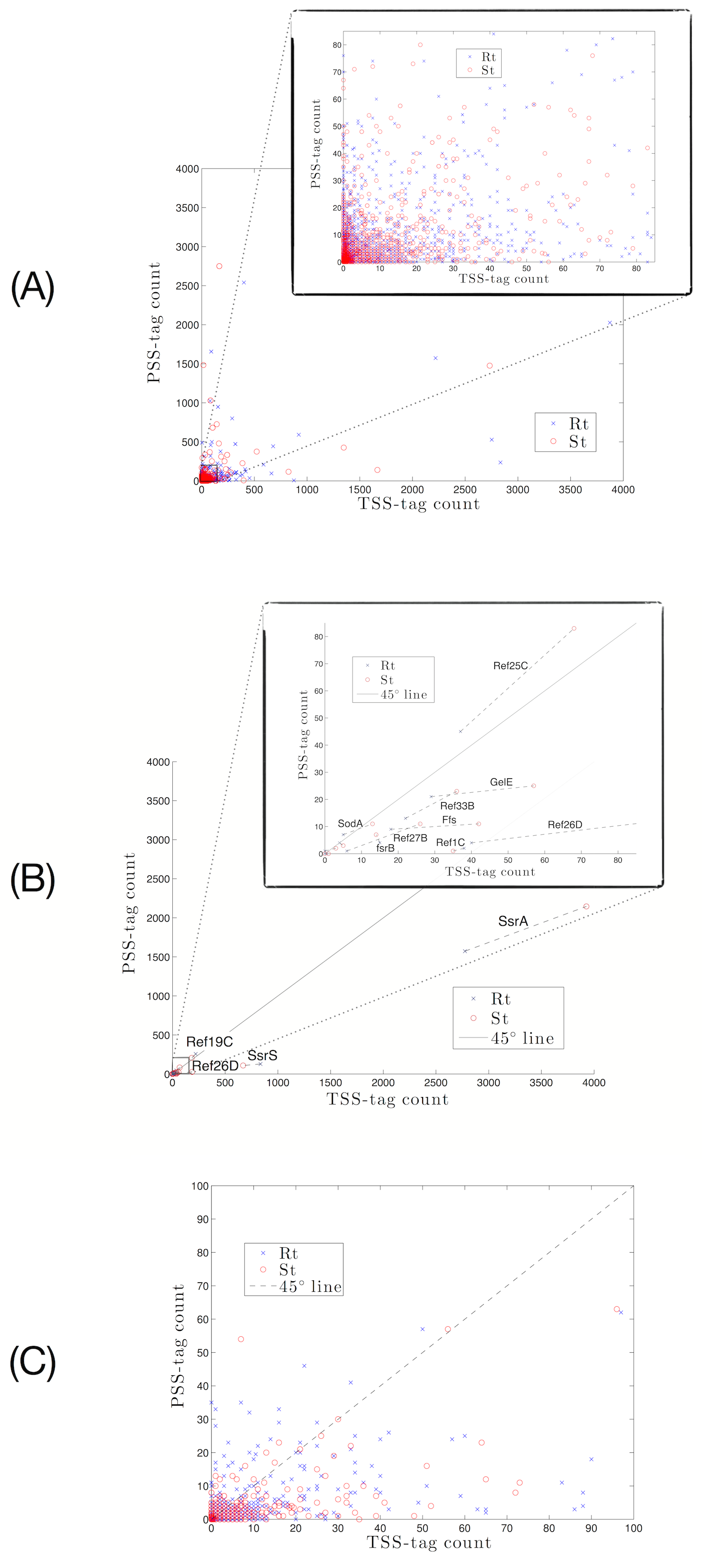}
  \caption{ {\bf Scatter plot showing TSS-tag counts versus PSS-tag counts.}
      {\bf (A)} at each position of the genome. {\bf (B)} at genomic locations within 2 bp of previously experimentally mapped transcription start sites. {\bf (C)} at genomic locations within 2 bp of 5' RNA edges of transcribed regions (see "Materials and Methods" and Section S3). About 80\% of 5' RNA ends predicted fall below the diagonal.}
      \end{figure}

\subsection*{Transcription start sites in \textit{E. faecalis}}
Within the area below 30\degree in figure 2A, we mapped a total of 559 TSSs on the V583 \textit{E. faecalis} genome, combining both St and Rt (Table SA and SB). A total of 327 TSSs were common to both transcriptomes. Among candidates classified as TSSs in St but not in Rt, 49 were classified as inconclusive due to a location between the 30\degree and 60\degree lines in figure 2A, 1 was classified as PSS and 27 were inconclusive due to a weak (TSS-tag + PSS-tag) signal (i.e. below 3.2x per million of reads aligned). For the corresponding candidates in the Rt conditions, those numbers are respectively, 36 between the 30\degree and 60\degree lines, 3 classified as PSSs and 116 were inconclusive due to low signal in St. 

\subsection*{Motif detection and promoter features in the \textit{E. faecalis} genome}
Up to date, less than 50 TSSs have been experimentally characterised in \textit{E. faecalis} \cite[and references therein]{FouquierdHerouel2011}. In order to better define promoter regions in this species, we took advantage of our comprehensive mapping and performed a blind search for common sequences nested in DNA regions preceding RNA extremities using the MEME suite \cite{Bailey2009}. By doing so, this search also enabled us to challenge our classification of 5' RNA ends based on the tagging method as presented in figure 2A. We defined four groups of DNA regions: two groups below the diagonal, one from 0\degree to 30\degree (called as TSSs), a second from 30\degree to 45\degree (called as undetermined, but expected to contain mainly TSSs), and two groups above the diagonal, one from 45\degree to 60\degree (called as undetermined but with a few TSSs, e.g. Ref25C, see Section S4), and a second group from 60\degree to 90\degree (called as PSSs). DNA sequences used as input for MEME and a detailed list of the motifs discovered are presented in table SC. For groups below 30$^\circ$, the analysis reveals motifs with strong statistical significance (E-values below $10^{-30}$ ) and consensus sequences: within the region $[-30\ldots0]$ and centered around position $-9.7\pm2.6$, we found GnTATAAT, the canonical -10 box; in the $[-40\ldots-20]$ region, the motif TTGACAA was found centered at $-31.5\pm2.3$, the canonical $-35$ box. The $-10$ box appears with a high frequency (83.5\%) and ends 5 to 9 bp from the 5' RNA ends mapped. The $-35$ box was found in 20.6\% of input sequences. At least 90\% of the sequences where a $-35$ box is detected also have a canonical $-10$ box. Boxes defined as $-10$ and $-35$ are spaced by a 16 to 22 bp long sequence. Thus, the most significant motifs discovered correspond to the canonical $-10$ (TATAAT) and $-35$ (TTGACA) sequences of promoters recognized by the vegetative RNA polymerase loaded with the transcription initiation factor SigA (RpoD, $\sigma^\textrm{\scriptsize A}$ or $\sigma^{70}$) in the most studied bacteria \textit{E. coli} and \textit{B. subtilis} \cite{Harley1987,Helmann1995}. The presence and the location of $-10$ and $-35$ boxes on DNA regions upstream 5' RNA ends falling in the area defined by the angle between 0\degree and 30\degree in figure 2A, reinforces our previous conclusion that these RNA extremities are TSSs. In line with this conclusion, for features with an angle between 0\degree and 45$^\circ$, $-10$ and $-35$ canonical boxes are still the most frequently found motif but the numbers fall to 80.8\% and 14.7\%, respectively, which indicates that the density of true TSSs is indeed higher for signal corresponding to a low angle in the plot ($\leq30$\degree). Within the two groups of sequences above 45\degree, the most significant motif discovered is AACG$\sfrac{\mathrm{A}}{\mathrm{T}}$A{$\sfrac{\mathrm{C}}{\mathrm{G}}$}{$\sfrac{\mathrm{A}}{\mathrm{G}}$} found in less than 10\% of sequences. To our knowledge, this purine-rich motif does not resemble any canonical sequence of bacterial promoter described previously. One might speculate that this sequence represents a frequent RNA motif targeted by an endoribonuclease, but further experiments will be required to confirm this hypothesis. Nonetheless, this observation reinforces our conclusion that the majority of TSSs do not locate above 45\degree in Figure 2A. \\
In addition to SigA, three other sigma factors have been predicted in \textit{E. faecalis} V583, SigH (Ef0049, the heat-shock factor), SigV (Ef3180, an "extracytoplasmic" factor) and SigN (Ef0782, a $\sigma^{54}$-like factor) \cite{Paulsen2003}. ORFs coding for SigH and SigV are not expressed in S and R growth conditions (Table SD and the ppRNome browser), hence we did not expect to find TSSs whose promoter regions would carry consensus sequences recognized by either one of these factors. In contrast, the \textit{sigN} encoding sequence is transcribed and we sought manually for the consensus sequence of SigN-dependent promoter ahead of TSSs mapped ($-$24/$-$12; TTGCCACNNNNNTTGCT) \cite{Hechard2001,Iyer2012,Buck01082000}. Only six corresponding locations were found across the whole genome: upstream ORFs coding for components of phosphor-sugar transfer systems (PTS), \textit{ef0019}, \textit{ef1012}, \textit{ef1017}, \textit{ef1954}, \textit{ef3210}, and \textit{fabF-2} coding for an enzyme involved in fatty acid and biotin metabolism. Out of those 6 locations, the TSS for \textit{ef1012} is detected and a tag signal below our selection threshold is found upstream of \textit{ef1017}.

\subsection*{Processing sites in \textit{E. faecalis}}
PSS-tags are found about 50\% more abundant than the number of total TSS-tags detected (Table SA and Section S2). In contrast to TSS-tags that appear with a discrete distribution at 5' edges or nested within transcribed regions, PSS-tags, in addition to co-localise with TSS-tags, tend also to spread out over RNA signals. Although we cannot rule out experimental RNA breaks, such a distribution of PSS-tags is expected as they label any type of 5' monophosphate RNA ends, including processing sites, degradation products and hydrolysed 5' triphosphate ends. To pinpoint major PSSs within the ppRNome, we only considered 5' ends located within the area delineated by the 60\degree angle in Figure 2A and above our acceptance threshold in both St and Rt. Ignoring rRNA and tRNA loci we mapped a total of 352 PSSs candidates (Table SE). 

Up to now, most of bacterial transcriptomic studies have focused on TSSs, RNA levels and the discovery of unannotated genes (e.g. \cite{Toledo-Arana2009,Sharma2010,Nicolas2012}). In addition to these aspects, the ppRNome visualizes RNA processing sites and shows that the "processed RNA landscape" is an important part of the total transcriptome that has often been overlooked. For example, the well-known ubiquitous sRNA RnpB, the ribozyme element of RNase P \cite{Frank1998}, provides an illustration of the information accessible in the ppRNome. We previously mapped the \textit{rnpB} TSS at location 1121939 in the \textit{E. faecalis} V583 chromosome \cite{FouquierdHerouel2011}, which is not detected by tagRNA-seq, most likely due to the higher amplification of the signal via the RACE-derivative method compared with the SOLiD procedure. The functional RnpB molecule, also termed M1, originates from a series of maturation processes conserved across the three domains of life that we may reasonably speculate to also operate in \textit{E. faecalis} due to the high degree of structural and functional conservation of RnpB \cite[ and references therein]{Griffiths-Jones2005,Li1998,Mann2003}. TagRNA-seq data enables us to map locations 1121951/-53 with high tag counts corresponding to PSSs (Tables SA, SE and the ppRNome browser). The RnpB upstream-most 5' end predicted in the Rfam database allocates a position at 1121944 in the chromosome \cite{Griffiths-Jones2005}, a location spaced by 4 and 7 nt from the TSS and PSS we have mapped, respectively. 

Further experiments will be necessary to shed light on the details of the processed transcriptome and its complex organization. Nevertheless, to our knowledge, this is the first study mapping PSSs at a global scale in bacteria.

\subsection*{Transcription start sites and Processing sites in \textit{E. coli}}

Unlike in \textit{E. faecalis}, transcription start sites in \textit{E. coli} have been relatively well studied and TSSs have been mapped with high accuracy for about 1000 ($\sim$ 23\%) of the about $4500$ ORFs in the \textit{E. coli} MG1655 genome \cite{Mendoza2009}. 

In order to challenge the tagRNA-seq method and our analysis, we applied the same procedure for the \emph{E. coli} transcriptome with a significance threshold set to 5 reads and we were able to retrieve 348 TSSs in the U00096.3 reference genome (Table SB). This lower number compared to \textit{E. faecalis} can be explained by the smaller number of reads obtained from this sequencing experiment (see section S2 in supplementary material)
while \textit{E. coli} has a genome about $40\%$ larger than \textit{E. faecalis}, out of which about 33\% (1.55 Mbp) appear to be transcribed (coverage higher than 2x). Out of those 348 TSSs, 98 (28\%) were found within 2 bp of a TSS mapped in \cite{Mendoza2009}.
This is in line with expectations given that 23\% of the \textit{E. coli} TSS were mapped in \cite{Mendoza2009}, and therefore supports the accuracy of tagRNA-seq.

The fraction of matching TSSs improves if a higher significance threshold is used, and can be brought up to 40\% of retrieved TSSs by using a threshold of 30 reads,
at the price of then retaining only 85 TSSs. This higher fraction is likely due to the fact that stronger transcription initiation sites are favoured regardless of the probing method used. We also observed that this fraction does not improve for angles below 30$^\circ$
while it worsens for angles around or above 45$^\circ$, confirming that 30$^\circ$ is a fair choice for TSS calling.

On the contrary, relatively few RNA processing sites have been mapped with single nucleotide accuracy in the \textit{E. coli} in standard growth conditions. Section S5 (Table S4) provides 16 examples of PSSs reported in the literature and how they appear in the ppRNome of \textit{E. coli}: i) eleven PSSs are clearly recovered and fall in the area above 60$^\circ$, albeit three carry tag counts below the chosen threshold of five reads; ii) five PSSs reported elsewhere are found within the area so-called "undetermined" (Figure 2). These examples support that the tagRNA-seq method enable us to map PSSs within the bacterial RNA landscape.

\subsection*{Non-annotated genes, small RNAs and particular transcriptional organization.}
Up to now, transcriptomic studies in \textit{E. faecalis} have used microarrays designed to examine expression of annotated ORFs \cite{Aakra2005,Abrantes2011,Makhzami2008,Mehmeti2011,Solheim2007,Vebo2009,Vebo2010,Vesic2013}, or custom-made tiling arrays containing a limited number of intergenic regions (IGRs) to search for sRNAs \cite{Shioya2011}. Although informative, these approaches provide partial information on the bacterial transcriptome, compared to RNA-seq methods \cite{Chao2012,Nicolas2012,Rasmussen2009,Toledo-Arana2009,Sharma2010,Sittka2008}. We took advantage of our comprehensive 5'RNA end mapping for a detailed transcriptional analysis, looking for previously non-annotated genes in the genome of \textit{E. faecalis} V583. Among other transcripts, sRNAs were primarily identified as stand-alone signals, whose length can be up to 500 nt long, located in "empty" regions (i.e. non-annotated regions), or transcripts antisense to annotated ORFs. In addition to the previously sRNAs identified \cite{FouquierdHerouel2011,Shioya2011}, we unveiled a total of 95 novel sRNAs (Figure 3 and Table SF). Considering our previous nomenclature \cite{FouquierdHerouel2011}, these new sRNAs were named from "Ref47" to "Ref120" when present in the chromosome, and for sRNAs encoded by plasmids pTEF1 and pTEF2, from "RefA1" to "RefA9" and "RefB1" to "RefB12", respectively. Five unnamed sRNAs reported in \cite{Shioya2011} were confirmed and named Ref77 (IGR \textit{ef1368}-\textit{1369} in the chromosome), RefA8 and RefA9 (IGR \textit{efa0080}-\textit{efa0081} in pTEF1), and RefB11 and RefB12 (IGR \textit{efb0062-63} in pTEF2), (Table SF). Over the last decade, sRNAs have been shown to ensure important regulatory functions and two major classes of sRNAs have been distinguished. Generally, regulatory RNAs not embedded in a transcriptional antisense organisation (stand-alone) modulate the activity of proteins or affect translation (up or down) by pairing to mRNAs; a class of sRNAs also named "\textit{trans}-acting sRNAs" \cite{Repoila2009,Waters2009}. Although not functionally characterised so far, many sRNAs found in \textit{E. faecalis} are likely \textit{trans}-acting regulators, e.g Ref50, Ref52, Ref72, Ref77, Ref79, Ref95, Ref102, RefA1, RefA4 (Table SF). Some sRNAs have been shown to carry a dual function since they can exert their regulatory role via different mechanisms or can also encode peptides \cite{Jorgensen2012,Loh2009,Livny2010,Sayed2012,Wadler2007}. Within the newly uncovered Ref sRNAs, some of them may encode for peptides as previously predicted for other sRNAs in \textit{E. faecalis} \cite{FouquierdHerouel2011}. AsRNAs is another remarkable category of sRNAs transcribed from the complementary DNA strand of genes, and thereby forming transcriptional antisense organisation. As regulatory consequence, the expression of an asRNA can impact the transcription initiation efficacy of the opposite gene (promoter interference), provoke premature arrests of transcription elongation, and/or modulate the translation and the stability of the cognate RNA \cite{Brantl2012,Georg2011,Sesto2013}. Many of the novel Ref sRNAs form antisense transcriptional organisation, e.g. Ref89 and RefB4 are antisense to sRNAs Ref90, RefB5, respectively; Ref94, Ref114 and Ref115 are antisense to transcripts bearing ORFs \textit{ef2025}, \textit{ef3087} and \textit{ef3088}, respectively (Table SF). 

\begin{figure*}[htbc]
  \centering
  \includegraphics[width=0.95\textwidth]{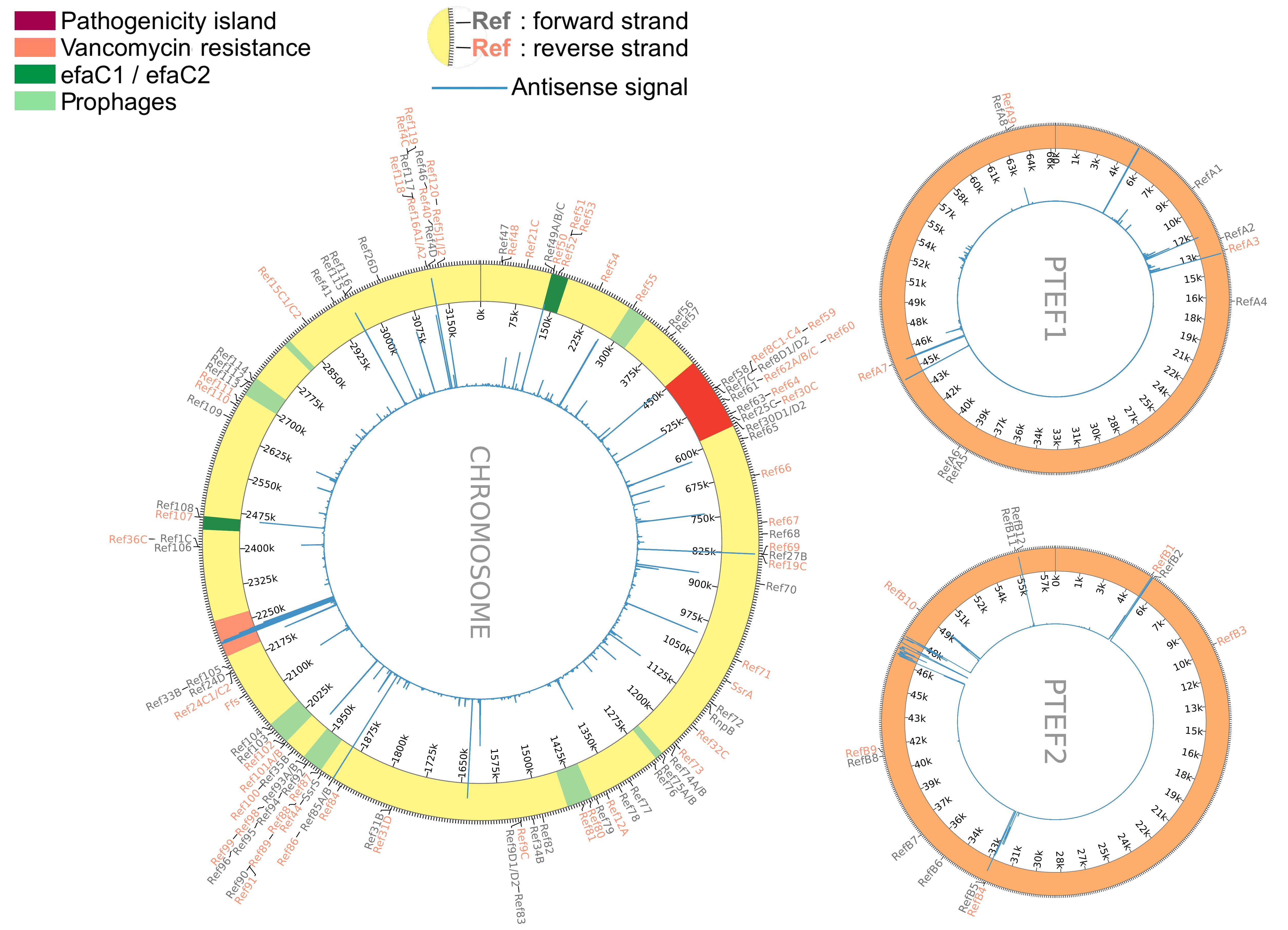}

  \caption{ {\bf Global view of sRNAs and antisense organisations currently known in \textit{E. feacalis} V583}
      (the chromosome and plasmids pTEF1 and pTEF2). The 95 new sRNAs discovered in the course of this work (Table SF) are emphasised in bold ('grey' on the forward strand; 'red' on the reverse strand). The inner plot visually describes the location and importance of antisense organisations detected (see the ppRNome browser for details). On the chromosome, the pathogenicity island (purple) and other mobiles genetic elements are annotated on the chromosome, i.e. efaC1/C2 (dark green), vancomycin resistance region (pink) and the six prophages (bright green) \cite{Lepage2006,Matos2013}. Antisense organisations are shown by vertical blue lines.}   
      \end{figure*}

Also, long 3' UTRs have been reported in several bacterial species and in a few cases their involvement in RNA-mediated regulations has been demonstrated \cite{Chao2012,Sittka2008}. Section S6, presents several cases found in the genome of V583. 

In addition, antisense transcriptional organisation also results from overlapping mRNAs and may involve coding sequences as well as 5'- or 3'UTRs (e.g. \cite{Nicolas2012,Rasmussen2009,Toledo-Arana2009,Sharma2010,Wurtzel2012}). For instance, the 5'UTR \textit{ef0282} (\textit{fabI}) overlaps the 5'UTR \textit{ef0283} (\textit{fab-F1}); ORFs \textit{ef0479} and \textit{ef0480} are embedded in a long opposite transcript originating \textapprox3,000 bp upstream; and the transcript that contains \textit{ef0522}-\textit{ef0523} in an operon is antisense to a transcript carrying \textit{ef0524}. Similar examples are observed on plasmids pTEF1 and pTEF2 (the ppRNome browser and Table SF). However, one of the most striking antisense organisation was found in the region spanning from \textit{ef2298} to \textit{ef2324} (Figure 4). It is well visible in each of the \textit{E. faecalis} transcriptomes, regardless of the growth condition, tagging or sequencing protocol used. It encompasses about 22 kbp on the chromosome and involves two transcribed regions of \textapprox 16 and 17 kbp long that overlap by \textapprox11.5 kbp. In the positive direction, the transcribed RNA originates 265 bp upstream of \textit{ef2304}, the unique predicted ORF contained within this 16 kbp long RNA and would code for a putative transcriptional regulator \cite{Paulsen2003}. In addition, this RNA is antisense to \textit{ef2312} and \textit{ef2314} that code for the DNA topoisomerase III (TopB-2) and a putative bacteriocin, respectively. On the minus direction, the second RNA originates \textapprox225 bp upstream of \textit{ef2308}, and carries \textit{ef2298} and \textit{ef2299} (Figure 4). These later ORFs encode for the two component regulatory system VanRB/SB, a vital element for \textit{E. faecalis} V583 to resist to vancomycin, a major clinical antibiotic against Gram positive infections \cite{Huycke1998,Arias2012}. Experimental validations will be required but it is tempting to speculate that this antisense regulation may control vancomycin resistance in \textit{E. faecalis} V583. 

\begin{figure*}[htb]
  \centering
  \includegraphics[width=0.95\textwidth]{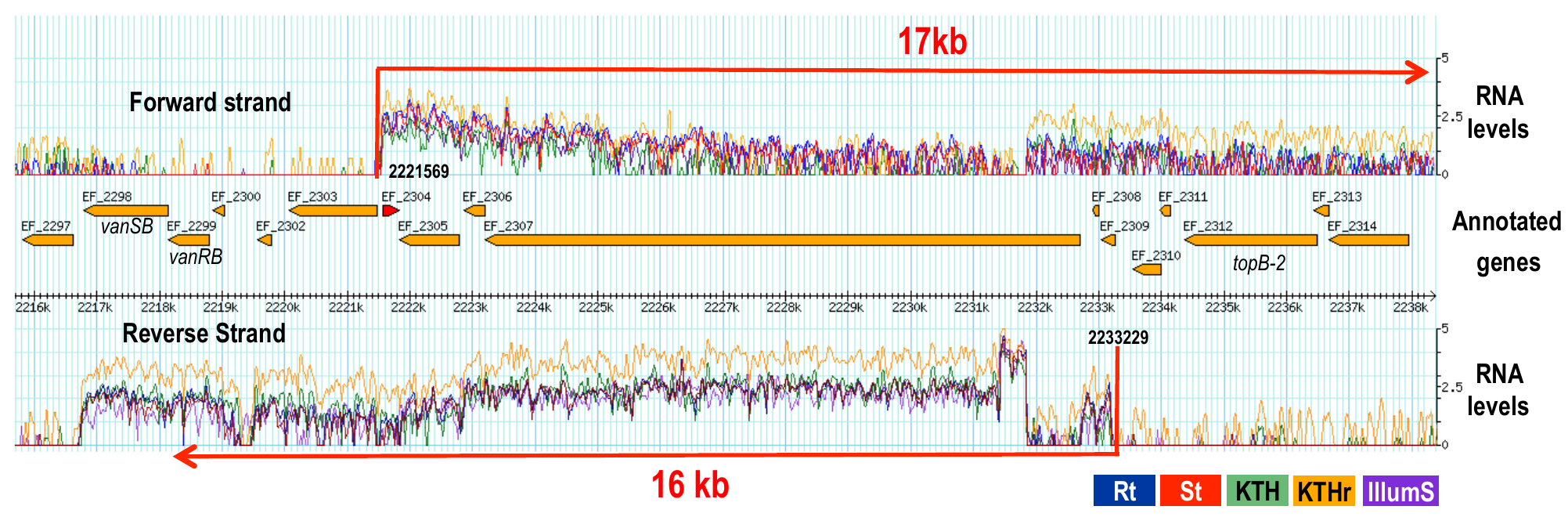}
  \caption{ {\bf Long antisense organization in the chromosome of \textit{E. faecalis} V583.}
      The transcriptional antisense organization encompasses 22 kbp. Note that only a single ORF, \textit{ef2304}, is predicted in the transcript originated from the positive DNA strand at coordinate 2221569. ORFs, \textit{ef2312} and \textit{ef2314} encode the DNA topoisomerase III and a putative bacteriocin; they are not transcribed in the growth conditions used but are "covered" by the antisense RNA (17 kb). vanRB and vanSB, encoding the regulatory two-component system of the vancomycin resistance locus, are contained at the end of the transcript originated at coordinate 2233229 from the minus DNA strand. The two antisense RNAs overlap by 11.6 kb. Coordinates mapped for TSSs of each corresponding transcript are noted at the corresponding location. The color boxes denoted by "Rt", "St", "KTH", "KTHr" and "IlluminaSt" applied to RNA levels of corresponding transcriptomes. RNA levels shown are normalized. }
      \end{figure*}

\section*{Conclusion}
In this work we have introduced a new method to distinguish primary and processed RNAs and achieved the first RNA-seq transcriptome of \textit{E. faecalis}. The discovery of numerous sRNAs and antisense organisations in \textit{E. faecalis} transcriptomes highlights, as like in many other species, the importance of RNA-dependent regulatory processes. The association of the RNA-seq method with the differential labelling of 5' RNA ends, enabled us to provide the two "faces" of a bacterial RNA landscape, i.e. the ppRNome. We mapped 559 TSSs and predicted promoter motifs at the genome-wide scale in a species where less than 50 were previously known, and  352 major PSSs, providing a first view of a bacterial processed RNA landscape. As TSS- and PSS-tags hallmark transcription initiation and processing, the next step in the exploitation of the ppRNome will be to perform quantitative studies in order to pinpoint the contribution of RNA synthesis and RNA stability in gene expression reprogramming accompanying physiological adaptation. This study constitutes a significant advance in the understanding of the organisation and the expression of the genetic information of the human pathogen \textit{E. faecalis}, and a key improvement of the functional analysis of bacterial transcriptomes.

\section*{Materials and Methods}

\subsection*{Bacterial growth and RNA preparation}
\textit{E. faecalis} V583 (VE14002 in our laboratory collection) was grown in brain-heart infusion (BHI) medium at 37$^\circ$C in static (S) or respiratory (R) conditions as described in \cite{FouquierdHerouel2011}. In the course of this work, we discovered that our laboratory strain did not contain the plasmid pTEF3 \cite{Paulsen2003}; although we used the appellation of "V583" throughout the text, data presented are those obtained for our strain VE14002. Total RNAs were prepared from bacterial cultures grown to an optical density (OD600) ranging between 0.7 and 0.85, as previously described \cite{FouquierdHerouel2011}.

\textit{E. coli} strain MG1655 was grown in LB medium at 37$^\circ$C under agitation (200 rpm) until an OD600 of 0.5. Bacteria were pelleted and total RNA prepared as previously described \cite{FouquierdHerouel2011}.

\subsection*{RNA tagging and sequencing}
5' RNA ends were differentially labeled with two short and different RNA oligonucleotides (tags), (\cite{FouquierdHerouel2011} and Section S1). Briefly, primary transcripts contain 5' ends with a triphosphate group which is brought by the first nucleotide triphosphate used by the RNA polymerase to initiate RNA synthesis at TSSs. In contrast, RNA processing events generate, at cleavage sites (PSSs), 5'ends with monophosphate groups. RNAs with PSS and hydrolyzed 5'-triphosphate RNA ends were tagged by a first ligation step with the PSS-RNA adaptor (PSS-tag). Subsequently, RNAs were treated with the tobacco alkaline phosphatase (TAP) to transform triphosphate groups into monophosphate groups and were then tagged by a second ligation step with a TSS-RNA adaptor (TSS-tag). TSS- and PSS-tag sequences were adapted to RNA-seq in such a way that they cannot be mistaken with any regions of the V583 reference genome (Section S1).
Two RNA libraries were obtained from total RNAs prepared from \textit{E. faecalis} grown in S and R conditions. They were tagged with PSS- and TSS-tags according to our 5' RNA end discriminative method, treated according to SOLiD manufacturer's protocols for sequencing (Applied Biosystems, Life Technologies Corporation), and sequenced on a SOLiD 5500 platform (MetaGenoPolis, INRA, France). The corresponding transcriptomes were named "St" and "Rt", respectively (Table SA).

Additionally, as control, three other RNA libraries prepared from two independent growths in S conditions were sequenced. 
In one experiment, the bacterial culture was grown at the Karolinska Institute, Sweden, as previously described in \cite{FouquierdHerouel2011}, and sequenced on a SOLiD v3 platform (Viiki, Finland). Two libraries, denoted as "KTHr" and "KTH" respectively, were prepared from this experiment. For one of them, ribosomal RNAs (rRNAs) were removed using Ambion MICROBExpress Bacterial mRNA Enrichment Kit; in the other, rRNAs were retained. 
In the second S growth culture, bacteria were grown at INRA as described in here and sequenced on a Hi-seq platform (IMAGiF, CNRS, France) following the Illumina Trueseq protocol, resulting in the "IlluminaSt" transcriptome. 

A single RNA sample was prepared from \textit{E. coli} total RNA, tagged using the same RNA adaptors (TSS- and PSS-tags) and sequenced on the SOLiD Wildfire platform (MetaGenomPolis, INRA, France). The resulting transcriptome was named "Coli".

\subsection*{Alignment and coverage}
Reads were aligned to the \textit{E. faecalis} v583 and \textit{E. coli} K12 substrain MG1655 reference genomes  (respectively GenBank Accession IDs [GenBank:AE016830.1] (chromosome), [GenBank:AE016831.1] (pTEF2), and [GenBank:AE016833.1] (pTEF1) ;  [GenBank:U00096.3] ) using Bowtie 1.0.0 \cite{Langmead2009} with default options, but allowing for multiple matches (\texttt{-a best} command line option). The coverage is calculated by counting the number of reads mapped at each position on the genome for each strand. In the case of multiple matches, the number of matches correspondingly divides the contribution to the read count. In cases mentioned explicitly in the text where repeated regions are excluded from the analysis; those multiply matched reads are ignored in the count. Similarly, when rRNAs and tRNAs are excluded, we impose a zero coverage over the corresponding regions. In order to reduce the effect of fragment bias (reads that are not uniformly distributed within the transcripts they represent) \cite{Roberts2011}, we defined a quantity called "coverage density" similar to the coverage, except that reads mapped so that they start at the same genomic position are counted only once. The resulting signal is thus less sensitive to the specific amplification of the different fragments at the cost of losing dynamic range. The coverage density signal has the useful and exploitable feature that edges of an expressed region are always staircase-shaped. We use the coverage density signal as a means to predict transcript edges from our RNA-seq data. Section S2 compiles the raw sequencing output for the various transcriptomes performed. 

\subsection*{Gene expression level}
We calculated gene expression levels of annotated genes of the \textit{E. faecalis} genome and performed differential expression analysis between R and S growth condition using Cuffdiff from the Cufflinks suite v2.1.1 \cite{Trapnell2010}. Cuffdiff was run on the Bowtie output files with the command line option \texttt{-u --library-type fr-secondstrand} using the genome of \textit{E. faecalis} V583 and its annotation. Regions corresponding to rRNA in the annotation were masked using the \texttt{-M} option. Results from the analysis are available in section S7 and table SD.

\subsection*{Predictions of transcription start sites}
Starting from the coverage density signal, we developed an iterative algorithm to detect transcribed regions, filtering out signals of low quality originating from sequencing errors or misalignments. The algorithm is inspired by the edge thinning operation in image processing \cite{Davies1981}. All regions where the signal is greater than a given but arbitrary confidence threshold are marked as "strong" signal. The signal in the immediate vicinity of this strong signal is recursively annexed to the strong signal region. All signals not marked as strong are discarded (Section S3). The algorithm discriminates low signals within transcribed regions and eliminates those likely caused by noise. The orientation of the aligned reads and the edges of signals enable us to assign TSSs. 

\subsection*{Detection of transcription starts and processing sites using 5' tags}
The addition of tags allows to readily map 5' ends of RNA molecules and to discriminate primary transcripts (ligated to TSS-tags) from processed transcripts (ligated to PSS-tags). Prior to alignment, reads are sorted according to tag sequences or their absence and, when present, tag sequences are removed from reads, leaving only sequences from bacterial RNAs. Both operations are performed simultaneously with Flexbar --- Flexible Barcode and adapter removal for sequencing platforms --- v2.4 \cite{Dodt2012} allowing for up to two mismatches in the 13 nt of the tags (command line parameters \verb!"--barcode-trim-end LEFT! \verb!--barcode-threshold 1.6 --barcode-! \verb!-unassigned --barcode-min-overlap 9! \verb!--min-ead-length 35"!). After alignment, reads with tags are classified into TSS or PSS candidates according to the rules described in the Results and Discussion section "5' tagging of RNA ends: analysis and interpretation". 

For transcriptomes obtained from the Illumina Hi-seq and SOLiD Wildfire, standard removal of 3' sequencing adapters is performed using Flexbar in an additional preprocessing step. This step is not needed for SOLiD v3 and 5500 where the insert size is typically much longer than read length \cite{Innocenti2013}.  

As a 5' RNA end can be tagged by both TSS-tag and PSS-tag sequences (see below), we considered a 5' end to be present in the sequenced RNA population when at a given location, the sum of TSS-tags and PSS-tags have counts of at least 3.2x per million of reads aligned in \textit{E. faecalis} transcriptomes. Such a threshold corresponds to a total of \textapprox5 tags (TSS- + PSS-tags) detected at the concerned position for the St transcriptome and \textapprox7  tags for Rt. For the "Coli" transcriptome, this threshold was kept at 5 tags, or 4.06x per million read aligned. 

On one hand, a careful examination of the transcriptomes reveals many instances of one or two isolated reads in isolated positions. It is reasonable to assume that many of these are noise and thus setting the limit above this level was chosen to eliminate them. On the other hand, as described in details in section S8, there seems to be no natural threshold in the data and a lower threshold simply leads to more candidates. The threshold values given above are simply one reasonable choice. 

It is well known that transcription initiation at a transcription start site is not always initiated with single nucleotide accuracy \cite{Sharma2010,23497287,Cortes:uo,Morton01072014}. To take this into account, when locations distant by 4 bp or less from each other have mapped reads with TSS-tags and at least one of them is classified as a TSS candidate, the multiple tag signals are grouped in a single region that encompasses all those locations (Tables SA, SB). The location with the highest tag signal is taken to be the most probable location of the TSS, and the total tag signal for the region is taken as the sum of the signals at all locations in the group. Although the length of such a "TSS region" can reach 6 bp or more in rare cases, many TSSs are detected with single nucleotide resolution (Figures S9a and S9b in section S9). Furthermore, an analysis of the average signal around retrieved TSSs shows that most of the signal concentrate within a region of $\pm$ 2 bp around the most probable location (Figure S9c in section S9). 

As much less is known about the accuracy of the different ribonucleases, PSSs are reported in the text as point location on the genome and neighbouring nucleotides with tag signal classified as PSSs are counted as different PSS sites.   

\subsection*{Motif detection}
We performed unbiased \textit{de novo} motif search using MEME v4.9.1 \cite{Bailey2009} upstream of genomic positions with TSSs and PSSs. The search was limited to the 10 most significant motifs with a width between 4 and 8 bp (command line arguments -nmotifs 10 -minw 4 -maxw 8). Short DNA sequences were extracted from the reference genome between [20 to 40 bp] and [0 to 30 bp] upstream of locations of interest and classified according to their ratios of PSS to TSS tag signals (as described in the Results and Discussion section "5' tagging of RNA ends: analysis and interpretation"). Those sequences were used as input to MEME without any filtering. When the feature was a TSS region as defined in the previous section, the most probable location was used as reference position for the sequence. The list of input sequences and results of the analysis are available Table SC. 


\subsection*{Visualisation of Results}
All coverage information and tag signals resulting from our experiments and analysis can be visualised in a user-friendly and interactive manner online at the address \url{http://ebio.u-psud.fr/eBIO_BDD.php} (website named "The ppRNome browser"). The visualisation uses the Genome Browser (GBrowse) (section S10), \cite{Stein2002}. The data presented online are also available in a numerical format in tables SA and SD for the Rt and St transcriptomes.


\section*{Competing interests}
  The authors declare that they have no competing interests.

\section*{Author's contributions}
NI, AdH, FR and EA designed the research. NI, MG, AdH and EA developed and performed the computational analysis and visualization on website. AdH, CL, RB, FW, PS and FR performed bacterial manipulation, RNA extractions and tagRNA libraries preparation. SK performed the SOLiD tagRNA sequencing. NI, AdH, PS, PB, FR and EA analyzed data and wrote the paper. All authors read and approved the final manuscript.

\section*{Acknowledgements}
We thank the eBIO computing platform of Orsay, particularly C.~Drevet and C.~Toffano-Nioche, for hosting the "ppRNome browser". We thank Ingemar Ernberg at Karolinska Institutet for his hospitality by providing lab space and equipment for AdH, and V.~Cantoni for the suggestion to use edge detection algorithms. We are grateful to the 'CPE team', A.~Gruss and D.~Halpern for comments and discussions, and S.~Gaubert, L.~Girbal, T.~Esquerr\'e and M.~Cocaign-Bousquet for communicating results prior to publication. Many thanks to our colleagues, P.~Palcy, V.~Bourgogne, N.~Eberlin and P.~R\'egent for invaluable administrative, IT and equipment-related support.

This research was supported by the Swedish Science Council through grant 621-2012-2982 (E.A),
and by the Academy of Finland through Finland Distinguished Professor programme and the Center of
Excellence COIN (E.A), and grant ANR-12-BSV6-0008 (ReadRNA) from the Agence Nationale pour la Recherche.

\section*{List of abbreviations}
\ \\ 
\begin{tabular}{ll}
{\bf asRNA} & {antisense RNA} \\
{\bf BHI} & {brain-heart infusion} \\ 
{\bf dRNA-seq} & {differential RNA-seq} \\
{\bf IGR} & {Intergenic Region} \\
{\bf ORF} & {Open Reading Frame} \\
{\bf ppRNome} & {primary and processed }\\
& {transcriptome} \\ 
{\bf PSS} & {Processing Site} \\
{\bf PTS} & {phosphor-sugar transfer systems}\\
{\bf RNA-seq} & {RNA sequencing}\\
{\bf Sdh} & {serine-dehydratase}\\ 
{\bf sRNA} & {small RNA} \\
{\bf TSS} & {Transcription Start Site} \\
{\bf UTR} & {UnTranslated Region} \\
\end{tabular}

\onecolumn

\bibliographystyle{plainnat} 
\bibliography{Innocenti_et_al_2014}      

\clearpage

\section*{Additional Files}
  \subsection*{Additional file 1 --- Supplementary Material (PDF)}
    Supplementary text in Adobe Portable Document Format containing sections S1 to S10, as referred to in the article.
  \subsection*{Additional file 2 --- Supplementary Table SA (XLS)}
    Table in Open Document Format with tagRNAseq data.
  \subsection*{Additional file 3 --- Supplementary Table SB (XLS)}
     Table in Open Document Format with transcription start sites retrieved by tagRNA-seq.
  \subsection*{Additional file 4 --- Supplementary Table SC (XLS)}
      Table in Open Document Format with motif identification from MEME.
  \subsection*{Additional file 5 --- Supplementary Table SD (XLS)}
     Table in Open Document Format with gene expression levels and standard differential analysis using Cufflinks.
  \subsection*{Additional file 6 --- Supplementary Table SE (XLS)}
    Table in Open Document Format with reported PSS sites.
  \subsection*{Additional file 7 --- Supplementary Table SF (XLS)}
    Table in Open Document Format with novel transcripts.

\end{document}